\documentclass{aa}

\usepackage{graphicx}
\usepackage{txfonts}
\usepackage{amsmath}
\usepackage{tabularx}
\usepackage{longtable}

\usepackage{natbib}
\usepackage{hyperref}
\usepackage{xcolor}
\usepackage{braket}
\usepackage{multirow}
\usepackage{subfig}
\usepackage[utf8]{inputenc}
\usepackage{xspace}
\usepackage{lineno}
\hypersetup{
colorlinks=false, 
linkcolor=blue,
urlcolor=red, 
linktoc=all
}
\bibpunct{(}{)}{;}{a}{}{,}

\def\m87{M87$^*$\xspace}
\def\sgra{Sgr~A$^*$\xspace}

\usepackage{comment}
\begin{document}

\title{Constraining the jet base emission of \m87 with past and future Event Horizon Telescope observations}
\titlerunning{}

\author{Noemi La Bella
          \inst{1}
          \and
        Michael Janssen \inst{1} 
        \and Britton Jeter \inst{2,3}
        \and Hendrik M\"uller \inst{4}
        \and Bram Van de Berg \inst{1}
        \and Hung-Yi Pu \inst{5, 6, 7}
        \and Paul Tiede \inst{8,9}
\and Heino Falcke \inst{1} 
          }

\institute{
Department of Astrophysics, Institute for Mathematics, Astrophysics and Particle Physics (IMAPP), 
Radboud University, P.O. Box 9010, 6500 GL Nijmegen, The Netherlands\\
\email{n.labella@astro.ru.nl}
\and
Finnish Centre for Astronomy with ESO, University of Turku, FI-20014 Turun Yliopisto, Finland
\and
Aalto University Mets\"{a}hovi Radio Observatory, Mets\"{a}hovintie 114, FI-02540 Kylm\"{a}l\"{a}, Finland
\and
National Radio Astronomy Observatory, 1011 Lopezville Rd, Socorro, NM 87801, USA
\and Department of Physics, National Taiwan Normal University, No. 88,  Section 4, Tingzhou Road, Taipei 116, Taiwan, R.O.C.
\and Centre of Astronomy and Gravitation, National Taiwan Normal University, No. 88,  Section 4, Tingzhou Road, Taipei 116, Taiwan, R.O.C.
\and Institute of Astronomy and Astrophysics, Academia Sinica, 11F of Astronomy-Mathematics Building, AS/NTU No. 1, Sec. 4, Roosevelt Rd., Taipei 10617, Taiwan,
R.O.C.
\and Black Hole Initiative at Harvard University, 20 Garden Street, Cambridge, MA 02138, USA
\and Center for Astrophysics | Harvard \& Smithsonian, 60 Garden
Street, Cambridge, MA 02138, USA
}

\abstract
{}
{We investigate the detectability of the jet base of \m87 at Event Horizon Telescope (EHT) observing frequencies. Although M87 is known to host a prominent relativistic jet, detecting jet emission close to the black hole at horizon scales remains challenging. Our goal is to determine the minimum jet intensity that can be reliably detected with the recent EHT array configurations.}
{We use synthetic EHT data generated for three array configurations corresponding to the 2021 and 2022 observing campaigns and to a near-future EHT campaign. As input models, we employ semi-analytic accretion–jet models in which the jet emission can be tuned independently of the accretion flow. The synthetic datasets are reconstructed using multiple imaging approaches, including regularized maximum likelihood and Bayesian imaging techniques. Jet detectability is assessed through flux density recovery, image fidelity, and uncertainty maps.}
{We find that jet detectability strongly depends on the jet intensity, the array configuration, and the imaging methodology. Using our analysis, we determine a lower limit on the jet intensity that can be reliably recovered. The 2022 EHT array configuration represents a significant improvement over earlier arrays, enabling a more robust reconstruction of faint jet features.}
{Our results indicate that the current EHT array is already sensitive to weak jet emission at horizon scales in \m87. The improved short-baseline coverage introduced in 2022 makes faint inner jet features more easily detectable. If the inner jet contributes a significant fraction of the unresolved compact flux, it should become visible in post-2021 observations. On the other hand, if no clear jet signature is found, this would suggest that the horizon-scale jet contributes only a small part of the compact emission. The continued expansion of the EHT will further improve our ability to detect such jet emission in \m87 .}

\maketitle
\nolinenumbers
\section{Introduction}

The nearby Virgo Cluster hosts one of the closest known Active Galactic Nucleus (AGN), located in the galaxy M87. This AGN has long attracted interest for its central supermassive black hole (SMBH), known as \m87, and for its relativistic jet \citep[e.g.][]{Abramowski2012, Giroletti2012,Algaba2021,Hada2024}. The Event Horizon Telescope Collaboration (EHTC) has released images of \m87's shadow based on data obtained during its 2017, 2018, 2021 observing campaigns, though additional observations have been made in subsequent years \citep{M87P1, M87P2, M87P3, M87P4, M87P5, M87P6, M87P8, M87P9, M872018}. The 2017 observations revealed a black hole shadow \citep{Falcke2000} with an angular diameter of $\theta_d = 42 \pm 3 \rm{\mu as}$, consistent with a black hole mass of $6.5 \times 10^9 M_{\odot}$ .
While the 2018 EHT observations reproduced the ring diameter observed in 2017, they revealed a notable change in the azimuthal brightness distribution, with the peak emission shifting by approximately $30^\circ$ counterclockwise. More recently, the EHTC presented a joint analysis of total intensity and polarized images of \m87 from 2017, 2018, and 2021, finding that the total intensity brightness position angle in 2021 is consistent with that observed in 2018, and that the polarized structure exhibits year-to-year variability \citep[see][for more details]{M872025}. These findings highlight the importance of upcoming multifrequency EHT campaigns and enhanced array observations to further refine our understanding of the \m87 accretion flow and jet-launching region.

M87 is known to host a powerful relativistic jet, whose emission on horizon scales remains only weakly constrained by current EHT observations.
The jet exhibits a non-thermal spectrum from radio to $\gamma$ rays that is consistent with synchrotron emission, steepening from the radio through the optical/UV. At sub-millimeter frequencies, the spectrum shows a "sub-mm bump" (in analogy to Sgr~A*, see \citealp{FalckeGossMatsuo1998} and references therein), indicating a transition to optically thin emission close to the black hole \citep{2013Doi}.
The relativistic jet has been detected across the electromagnetic spectrum over the past decades \citep[e.g.][]{Biretta1999,Owen2000, 2002Marshall, Gebhardt2011,Perlman2011, deGasperin2012,Algaba2021,Algaba2024,2025Roder}.
At 3 mm wavelength, the Global mm-VLBI Array (GMVA) and Very Long Baseline Array (VLBA) have resolved the inner jet down to tens of Schwarzschild radii. These observations reveal a wide opening angle of $ \sim 60^{\circ}$ at the jet base, which gradually re-collimates into a narrower conical structure on larger scales \citep[e.g.][]{Junor1999, Hada2011, Asada2014, Kim2018, Walker2018, Janssen2019, Okino2022, Yang2024}. 
General relativistic magnetohydrodynamic (GRMHD) simulations of magnetized accretion flows \citep{Narayan2003,Moscibrodzka2016} show that relativistic outflows can be launched through magnetic processes, including the extraction of black hole spin energy via the Blandford–Znajek mechanism \citep{Blandford1977} as well as disk-driven winds powered by accretion through the Blandford–Payne mechanism \citep{Blandford1982}. The observed wide opening angle and gradual parabolic collimation in \m87 agree with these models. 

Recent 86 GHz GMVA+ALMA+GLT observations revealed a spatially resolved ring-like central structure together with an inner jet showing clear limb brightening and extending to the immediate vicinity of the core \citep{Lu2023}. Subsequent re-imaging analyses confirmed the robustness of this ring+jet morphology, while also showing that finer details of the innermost emission, such as the prominence of a possible central spine, depend on the imaging method and angular resolution \citep{Kim2025}. These results indicate that the inner jet in  M87 at 86 GHz is morphologically more complex than a smooth, centrally filled outflow and may contain localized bright features superimposed on a broader emission component.

The EHT, operating at 230~GHz with an angular resolution of $\sim20~\mu\rm{as}$, probes spatial scales of $\sim10 R_{\rm s}$ in \m87 that correspond to the optically thin region close to the black hole. At this resolution, the array images the ring-like structure surrounding the black hole shadow and the innermost jet base, consistent with the jet-launching region predicted by GRMHD simulations. Early analyses of the 2017 EHT data reported evidence for faint emission offset from the black hole shadow that may be associated with the jet base \citep{Broderick2022,Arras2022,Carilli2022}. The GRMHD models that describe the 2017 EHT data best are those with strong jet emission \citep{Janssen2025c}. However, the absence of short and intermediate baselines in the 2017 and 2018 EHT array prevented robust detection. In 2021, new shorter baselines were provided by the NOrthern Extended Millimeter Array (NOEMA) in France and the Kitt Peak (KP) telescope in Arizona. Using this configuration, \cite{M872025} reported jet emission outside the ring. In particular, \citet{Saurabh2026} investigated hints of faint extended emission near the base of the jet, modeled as a Gaussian component with a preferred location roughly $300 \,\mu{\rm as}$ west and $100 \,\mu{\rm as}$ south of the ring center, improving the closure phase fits on intermediate baselines. \cite{Georgiev2026} further demonstrated that residual offsets in triangles that involve ALMA and APEX can be explained by an additional $\sim1$ mas component located northwest of the ring, consistent with the larger-scale jet. Simulations incorporating the Korean VLBI Network (KVN) to the EHT \citep{Cho2025} further show that short baselines at multiple frequencies (86–230~GHz) significantly improve recovery of both the jet and the ring structure.

Motivated by these advances, we assess the limitations of past and upcoming EHT configurations in detecting the \m87 jet base. Our tests use synthetic observations from multiple years, semi-analytic models with varying jet intensity, and different imaging pipelines that include recent advanced Bayesian and Regularized Maximum Likelihood (RML) approaches previously used on real EHT data.

The paper is organized as follows. Sect.~\ref{Sec:array} describes the EHT observing campaigns and the participating telescopes for the years analyzed. Sect.~\ref{Sec:Methods} presents the input models and imaging methods adopted in this work. In Sect.~\ref{Sec:results}, we first fix a single observing campaign and investigate how the detectability of jet emission changes as the jet intensity is progressively reduced in order to identify a lower detection limit; then we explore how this limit varies across the different EHT observing campaigns considered.
Finally, Sect.~\ref{Sec:conclusions} summarizes our findings and discusses prospects for future EHT observations and jet detectability.

\begin{figure*}[t]
    \centering
 \includegraphics[width=\linewidth]{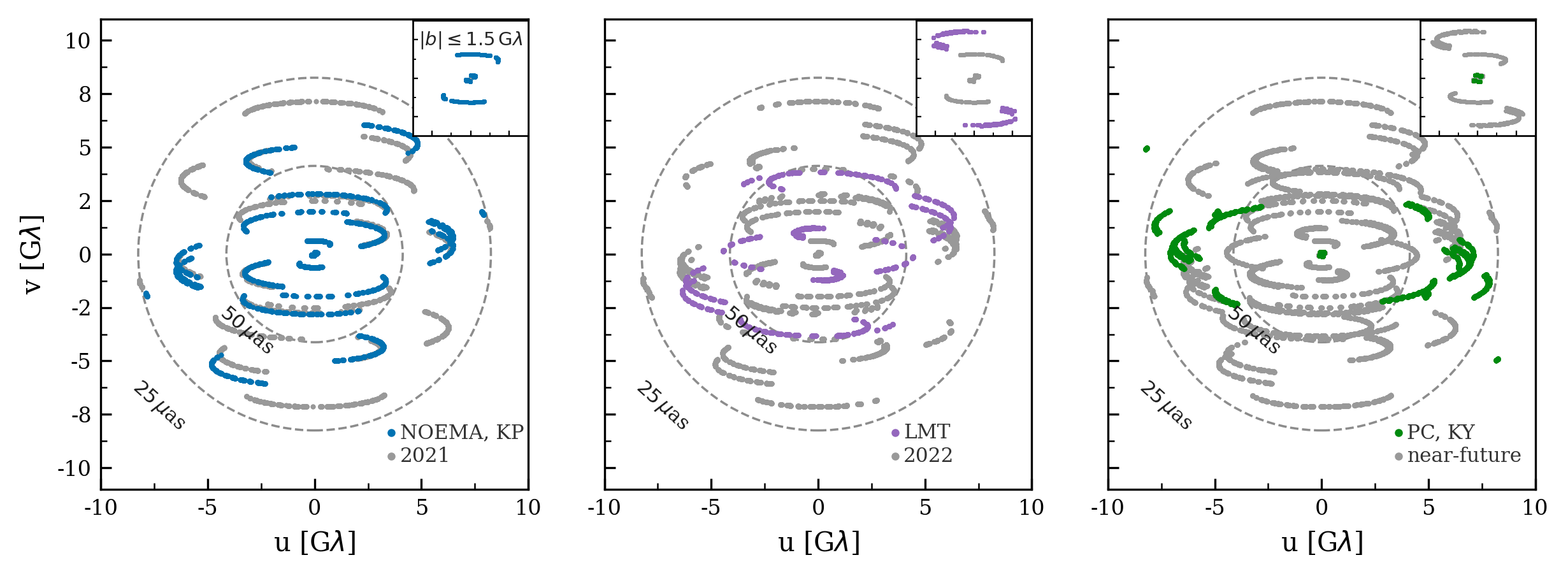}
    \caption{Projected $(u,v)$ coverage at 230~GHz for three representative EHT configurations (2021, 2022, and near-future). Colored points indicate newly introduced baselines in each configuration. Dashed circles correspond to angular scales of 25 and 50 $\mu$as. Insets highlight the short-baseline coverage ($|b| \leq 1.5$ G$\lambda$), which remains relatively limited.} \label{fig:uvplots}
\end{figure*}
\section {EHT Development and campaigns}
\label{Sec:array}

The EHT has undergone continuous expansion and technical upgrades since the initial observations of 2017. At that time, the array was composed of eight telescopes: the Atacama Large Millimeter/submillimeter Array (ALMA) and the Atacama Pathfinder Experiment (APEX) in Chile; the IRAM 30-m Telescope at Pico Veleta (PV) in Spain; the Submillimeter Array (SMA) and the James Clerk Maxwell Telescope (JCMT) in Hawaii; the Submillimeter Telescope (SMT) in Arizona; the Large Millimeter Telescope Alfonso Serrano (LMT) in Mexico; and the South Pole Telescope (SPT) in Antarctica.
In 2018, the Greenland Telescope (GLT) joined, the observing bandwidth doubled (from $\sim$4~GHz to $\sim$8~GHz, except at GLT for that year), implemented as four bands centered at sky frequencies of 213.1~GHz (band 1), 215.1~GHz (band 2), 227.1 GHz (band 3), and 229.1 GHz (band 4). In addition, the effective aperture of LMT increased by $\sim 50 \%$.  In 2021, the array expanded further with the addition of NOEMA and the KP telescope, although the LMT was unable to participate due to logistical issues. By 2022, full participation of the core stations was achieved. All major campaigns to date have been conducted at 230~GHz (1.3 mm), while most stations are also being prepared for observations at 345~GHz (0.87 mm), enabling future global campaigns at this frequency. Since 2024, the inclusion of the KVN station Yonsei (KY) has enhanced both long- and short-baseline coverage, while the PyeongChang station (PC) is expected to participate in the next campaigns. Looking ahead, further expansion will include the Africa Millimeter Telescope (AMT, Namibia) and the Canary Island Telescope (CNI, Spain), which will provide important additional coverage of the southern sky \citep{LaBella2023}, as well as future extensions with the ngEHT array \citep{Doeleman2019}. 

Despite these advances, reconstructing the diffuse inner jet structure of \m87 remains challenging due to the lack of short baselines at $(u,v)$ distances below $\sim$1 $G \lambda$. Fig.~\ref{fig:uvplots} shows the projected $(u,v)$ coverage at 230~GHz for three representative EHT configurations considered in this work.
In the first panel, we show the 2021 $(u,v)$ coverage. The addition of short and intermediate spacings from NOEMA and KP enhances sensitivity to extended structure. The PV–NOEMA baseline is $\sim  1100$\,km, yielding a nominal fringe spacing of $\sim 250\,\mu \rm{as}$  at 230~GHz. The KP–SMT baseline is $\sim 100$\,km, corresponding to $\sim 2500\,\mu \rm{as}$. Although these new baselines are significantly shorter than those used in earlier campaigns, direct imaging of the extended jet base structure remains limited by sparse $(u,v)$ coverage and faint jet base emission.
The second panel represents the 2022 array. The inclusion of the LMT restores high-sensitivity mid-to-long baselines between North America and Europe/Chile, in particular reintroducing the LMT–SMT baseline, which in the 2017 and 2018 arrays was the shortest baseline between non–co-located sites. Although the short spacings remain sparse, the 2022 configuration provides for the first time a closed triangle of intermediate baselines (LMT–SMT–KP), enabling robust phase constraints across the angular scales relevant to the jet base. As a result, the array is more sensitive to both compact and jet emission compared to 2021. Nevertheless, the dedicated 2022 \m87 night was affected by poor weather, and some stations were unable to observe. We therefore used another \m87 night from the same campaign, which had fewer \m87 scans but full array participation. The 2022 configuration used here should be interpreted as a realistic, but not best-case, 2022 setup. For this reason, we simulated the near-future array, which includes the two additional KVN stations.
These three cases (2021, 2022, and near-future) were selected to investigate the lowest detectable jet emission in our models and illustrate the progressive filling of the $(u,v)$ plane compared to the sparser 2017 and 2018 coverage shown in Appendix~\ref{Sec:2017}.

\section{Methods}
\label{Sec:Methods}

\subsection{Semi-analytic models}

To enable flexible control over the relative flux contributions of the accretion and jet components, we employ semi-analytic models that explicitly incorporate both an accretion flow and a relativistic jet. The accretion component is implemented following the formalism described in \citet{PU2018}, which is based on \citet{2006Broderick}, while the jet component is constructed on the basis of the jet model developed in \citet{PU2020}. The jet model assumes a terminal Lorentz factor $\Gamma=10$, with the local Lorentz factor varying along the jet. At the projected axial distances $z = 100\,\mu$as and $z = 250\,\mu$as from the ring centroid, the Lorentz factor is approximately 
$\Gamma \approx 3$ and $\Gamma \approx 5$, respectively. The dimensionless black hole spin is chosen as $a=0.9$ for demonstration purposes. Note that this is likely opposite to the preferred spin direction in the actual source \citep{M87P5,M872018p2, Janssen2025c}. The goal of this project is to investigate the general reliability of detecting the extended emission, and not the detailed reconstruction of the precise jet structure from a particular semi-analytic model.  The emissions arising from the accretion flow and the jet are thermal synchrotron and non-thermal synchrotron radiation, respectively. Furthermore, our model images cover a $1\,\mathrm{mas}$ Field Of View (FOV) and are intended to represent resolved jet emission on the angular scales probed by the short and intermediate EHT baselines at 230 GHz, not the detailed near-horizon jet-launching region. Our analysis of the model jet emission is therefore mainly relevant on scales $z \gtrsim 100\,\mu\mathrm{as}$, comparable to the angular scales associated with the 230 GHz "jet base" emission discussed by \cite{Saurabh2026}.

To phenomenologically simulate various potential realizations of the scenarios considered, we construct exploratory models with differing relative contributions from the accretion flow and jet components. The relative ratios between these components are calibrated by varying the accretion and jet contributions through normalizations of the electron number density in the models.

The total flux densities of our semi-analytic models span 0.8, 1, and 3.6~Jy at 230~GHz. The accretion flow emission is kept relatively fixed, and the different total flux levels are obtained by mainly varying the jet emission. The lower-flux density models (0.8–1~Jy, both with accretion component  $\sim$ 0.6~Jy) correspond to more compact-dominated emission states similar to those commonly assumed in EHT imaging analyses \citep{M87P4,M872018}. The higher-flux density model (3.6~Jy, with accretion component $\sim$ 1.27~Jy) represents a jet-enhanced case in which a significant fraction of the emission is distributed over a more extended jet structure. 

Recent results of the 2021 EHT data suggest that the jet emission on sub-milliarcsecond scales is likely faint, with only weak deviations from pure ring models detected. Closure phase analyses favor a faint additional component of order $\sim60$ ~mJy near the ring, indicating that the jet emission on these scales is likely well below the Jy level \citep{Saurabh2026}. This makes the lower-flux models particularly relevant for assessing the detectability of weak horizon-scale jet emission with different EHT array configurations, while the strongest-jet model serves as a reference case in which the jet should be clearly detectable.

The jet axis in the main analysis is rotated on the sky to match the millimeter-scale jet orientation observed at 86~GHz \citep{Lu2023, Kim2025}. Models with different jet orientations were also explored in blind imaging tests (see Appendix \ref{Blind_test}) to assess the robustness of the reconstructions. 
\subsection{Synthetic images}
We generated synthetic EHT observations of our jet models using \textsc{Symba}  \citep{Roelofs2020} \footnote{
\url{https://bitbucket.org/M_Janssen/symba}}, an end-to-end VLBI simulation pipeline that incorporates atmospheric, instrumental, and calibration effects, producing datasets directly comparable to real EHT observations. \textsc{Symba} has previously been used to generate synthetic images of \m87 \citep{M87P1}, \sgra \citep{SgraP6}, to investigate jet structure in nearby radio galaxies such as Centaurus~A \citep{Janssen2021}, and, more recently, for large-scale inference studies based on millions of simulated EHT datasets \citep{Janssen2025a,Janssen2025b,Janssen2025c}.
For each model and array configuration, we simulated two independent datasets corresponding to the EHT band pairs, assigning each a single observing frequency equal to the mean central frequency of its two spectral windows ($\sim 214$~GHz for Bands 1–2 and $\sim 228 $~GHz for Bands 3–4). The resulting $b1$–$b2$ and $b3$–$b4$ datasets were produced for the 2021, 2022, and near-future campaigns, with realistic variations in schedule, atmospheric conditions, and station availability (see Appendix~\ref{Sec:SYMBA} for more details).

Synthetic data were reconstructed using both RML and Bayesian imaging approaches. At 230~GHz, the EHT provides sparse $(u,v)$ coverage and limited sensitivity to large angular scales, leading to degeneracies in image reconstruction and to the well-known problem of missing large-scale flux density.
In this regime, all imaging approaches become to some extent sensitive to method-dependent assumptions and user choices. For traditional deconvolution techniques such as CLEAN, for example, restricting components to predefined regions aligned with the expected jet axis can bias the reconstruction by forcing emission to appear along that direction.
We therefore adopt forward-modeling approaches that operate directly in the visibility domain, make prior assumptions explicit, and allow for residual station gain errors to be incorporated into the reconstruction. 
For RML imaging, we use \textsc{eht-imaging} \citep{Chael2016,Chael2018}, the EHT-standard imager employed since the first \m87 imaging analyses, which minimizes a weighted combination of data fidelity terms and image regularizers. Within the Bayesian framework, we use \textsc{Themis}, which has also been extensively applied in the \m87 imaging campaigns, modeling the source emission with a thin geometric crescent and adaptive spline raster and explicitly marginalizing over station gains \citep{Broderick2020, Broderick2020_raster, Broderick2022_PhotonRing, M872018}. We additionally employ the more recent \textsc{Resolve}  \citep{Resolve2023} and \textsc{Comrade} \citep{Tiede2022, Tiede2026_HIBI} frameworks, which also implement fully Bayesian image reconstructions using Gaussian-process or spline-based image priors and joint inference of source structure and instrument (gain) parameters. For the Bayesian methods, we present the mean reconstructed image from the inferred posterior distribution. These imaging approaches are standard methods used in EHT analyses and have been widely applied to \m87 and \sgra \citep[e.g.,][]{M87P1, M87P2, M87P3, M87P4, M87P5, M87P6, M87P8, M872018, M872025, SgrAP1, SgrAP2, SgrAP3}. Here, they are used as reconstruction tools, as our primary goal is to evaluate how different jet intensities and array configurations affect the detectability of jet emission rather than to compare imaging methods.
Moreover, the imaging pipelines were not simply applied with their standard settings used in previous EHT analyses. Instead, the reconstruction parameters, calibration strategies, and model priors were adjusted to improve sensitivity to diffuse jet emission in the presence of a dominant ring component. The specific settings adopted for each method are summarized in Appendix~\ref{sec:methods details}.

\begin{figure*}[h]
    \centering
 \includegraphics[width=\linewidth]{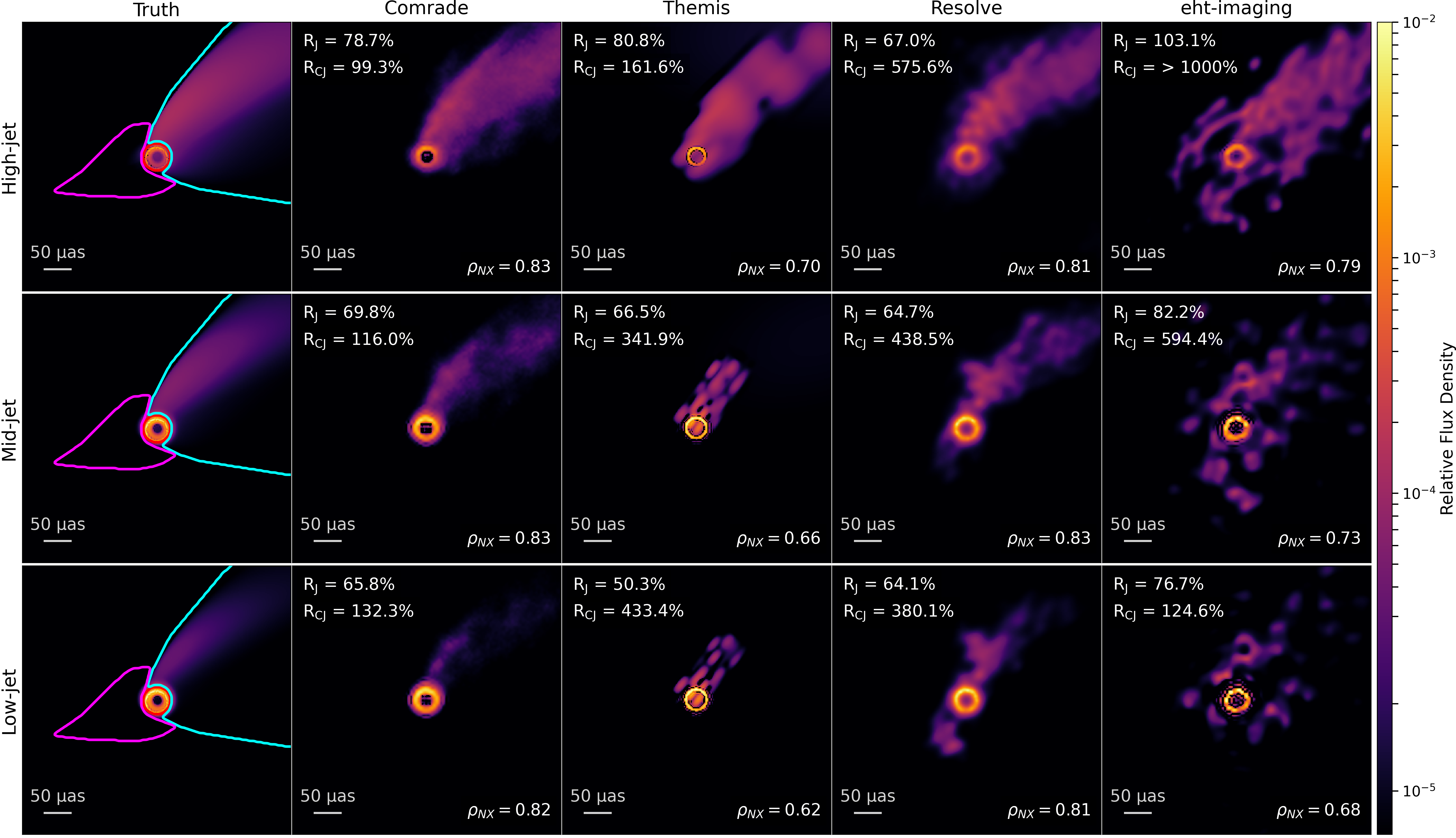}
    \caption{Jet recovery for the 2022 array configuration.
Rows show the high-, mid-, and low-jet truth models, while columns display the truth image and reconstructions obtained with the different imaging pipelines.
All images are aligned to the truth and normalized to a total flux density of 1~Jy.
The red, cyan, and magenta regions indicate the ring, jet, and counter-jet, respectively.
For each reconstruction, the recovered jet and counter-jet flux density ratios $R_{\rm J}$ and $R_{\rm CJ}$, as well as the normalized cross-correlation coefficient $\rho_{\rm NX}$, are reported in each panel.
The counter-jet is significant only in the high-jet model (see Appendix~\ref{sec:counterjet}) .
}\label{fig:2022jetrecovery}
\end{figure*}

\section{Results}
\label{Sec:results}
We present the results of the jet-detectability analysis at EHT frequencies. We first fix a single observing campaign and investigate how the detectability of jet emission changes as the jet intensity is progressively reduced in order to identify a lower detection limit. We then explore how this limit varies across the different EHT observing campaigns considered. The results shown here correspond to the 
 $b1$-$b2$ dataset; a comparison assessing band consistency with  $b3$-$b4$ is presented in Appendix~\ref{sec:Band_consistency}. 

\subsection{Jet detectability}
\label{sec:2022recovery}
To assess the jet detectability, we adopt the 2022 EHT array as our reference configuration. It provides strong baseline coverage through the reintroduction of the LMT, the availability of short baselines in Europe and Arizona, and for the first time includes an intermediate triangle of baselines (LMT–SMT–KP) sensitive to angular scales relevant to jet emission.
In Fig.~\ref{fig:2022jetrecovery} we show, for each input model, the reconstructed images obtained with the four imaging pipelines with the 2022 configuration. Similar plots for the 2021 and near-future arrays are shown in Fig.~\ref{fig:2021comparison} and Fig.~\ref{fig:nearfuture}, respectively. 
The rows show the three input models and the columns show the truth image (first column) and reconstructions (other columns).
For a fair comparison across imaging methods and input models,  we place all images on a common geometric and flux density scale. 
On the truth images, we define fixed regions for the ring (red), jet (cyan), and counter-jet (magenta) using the region file format of the \textsc{SAOImage DS9} image viewer, with the jet and counter-jet regions explicitly excluding the ring. Because the centroid of a ring mask can be biased by pixelization and brightness asymmetries, we refined the ring center of the truth image by searching for the center that maximizes the mean brightness within a ring-shaped annulus, with the inner and outer radii of this annulus derived directly from the ring region.
Each reconstructed image is then resampled to the truth pixel scale and aligned by shifting it so that its core position coincides with the truth ring center. After alignment, both the truth and reconstructed images are normalized to a total flux of 1~Jy. We then refer to the models as high-, mid-, and low-jet models. 

Flux density recovery is quantified through a pixel-by-pixel comparison between the reconstructed and truth images within the \textsc{DS9} regions defined on the truth image. The recovery values reported in the panels are defined as:
\begin{equation}
R_{\rm J} =
\left\langle
\frac{I_{\rm rec}}{I_{\rm truth}}
\right\rangle_{\rm J},
\qquad
R_{\rm CJ} =
\left\langle
\frac{I_{\rm rec}}{I_{\rm truth}}
\right\rangle_{\rm CJ},
\end{equation}
where $I_{\rm rec}$ and $I_{\rm truth}$ denote the reconstructed and truth pixel intensities, and the averages are taken over pixels in the jet and counter-jet regions, respectively. To avoid artificially large ratios from very faint pixels, we restrict the analysis to pixels with $I_{\rm truth} > 0.02 \, I_{\rm truth}^{\rm max}$, where $I_{\rm truth}^{\rm max}$ is the maximum pixel intensity of the truth image.
For completeness, we also report the integrated flux densities within the jet ($F_{\rm J}$), counter-jet ($F_{\rm CJ}$), and outside regions ($F_{\rm out}$). These are computed by first summing the pixel intensities within each region and then deriving the corresponding flux density fractions. The resulting values are listed in Tab.~\ref{tab:flux_recovery_2022} (and in Tabs.~\ref{tab:fluxes_2021} and \ref{tab:fluxes_nearfuture}). We note that $F_{\rm J}$, $F_{\rm CJ}$, and $F_{\rm out}$  are defined on the normalized 1 Jy images and therefore describe how the total flux is distributed among the selected regions, rather than the intrinsic flux density of the input models, which span 0.8, 1, and 3.6~Jy over the full FOV of 1~mas. In the panels, we additionally report the normalized cross-correlation coefficient $\rho_{\rm NX}$ between each reconstruction and the truth image as a global similarity metric. Because this metric is dominated by the bright ring emission, it primarily reflects the overall image fidelity rather than the detailed structure of the faint jet.
The detectability of the jet is therefore mainly quantified through the recovered jet flux density ratio $R_{\rm J}$, while the integrated flux density outside the defined regions ($F_{\rm out}$) serves as a diagnostic of misplaced emission.

In the high-jet  model, all imaging pipelines recover a large fraction of the jet emission, with typical values of $R_{\rm J}\sim 70-100 \%$ and $\rho_{\rm NX} \geq 0.8$. As the intrinsic jet brightness decreases, the recovered jet flux density ratio systematically decreases, with values of $\sim65 - 80\%$ for the mid-jet model and  $\sim50 - 70\%$ for the low-jet model.
In contrast, the counter-jet is significant only in the high-jet model, while in the mid- and low-jet models its true flux density is much weaker. As a result, $R_{\rm CJ}$ is often elevated and can exceed $100 \%$, as even small amounts of reconstructed emission translate into large relative values. This reflects the intrinsically weak counter-jet signal, consistent with previous VLBI studies of \m87\citep[e.g.][]{Ly2007, Walker2018}.

This effect is also visible in Tab.~\ref{tab:flux_recovery_2022}, where elevated values of $F_{\rm CJ}$ and $F_{\rm out}$ indicate misplaced flux density in the counter-jet and surrounding regions. The color scale adopted in Fig.~\ref{fig:2022jetrecovery}, optimized for the jet structure, further limits the visibility of the faint counter-jet emission in the mid- and low-jet models. A clearer view, together with additional details, is reported in Appendix~\ref{sec:counterjet}.
 
Among the tested imaging pipelines, \textsc{Comrade} provides the most conservative flux recovery, while also maintaining high $\rho_{\rm NX}$ and low $F_{\rm out}$, and avoiding extreme counter-jet recovery. For this reason, we adopt \textsc{Comrade} as a reference imaging pipeline for the campaign comparison analysis presented in the following section.
Based on the overall imaging quality, we did not consider jet models fainter than 0.8~Jy, as the imaging pipelines already exhibit clear difficulties in reliably recovering the jet at this level (see Appendix~\ref{Blind_test}).

An additional concern is whether the recovery of the faintest jet models is driven by the data or is instead influenced by prior knowledge of the jet direction. Indeed, a re-analysis of the 2017 EHT data reported a core–jet morphology without a dominant ring when deconvolution was restricted to predefined boxes aligned with the known jet axis \citep{Miyoshi2022}. Without these constraints, a reliable reconstruction of the jet structure and orientation from the 2017 data alone is not possible. To avoid a similar bias, we performed blind tests with different jet orientations that were not disclosed to the imagers (Appendix~\ref{Blind_test}). These tests show that the ring is robustly recovered even when the jet direction is unknown, while recovering the jet structure itself remains challenging, especially for the faintest models.

\begin{table}
\centering
\caption{Integrated flux densities for the 2022 EHT array.}
\label{tab:flux_recovery_2022}
\begin{tabular}{llccc}
\hline\hline
Model & Pipeline & $F_{\rm J}$ [Jy] & $F_{\rm CJ}$ [Jy] & $F_{\rm out}$ [Jy] \\
\hline
high-jet & Truth                & 0.5347 & 0.0075 & 0.0098 \\
                & \textsc{Comrade}     & 0.4423 & 0.0077 & 0.0600 \\
                & \textsc{Themis}      & 0.5636 & 0.0147 & 0.1062 \\
                & \textsc{Resolve}     & 0.4510 & 0.0330 & 0.1403 \\
                & \textsc{eht-imaging} & 0.4992 & 0.0582 & 0.1026 \\
\hline
mid-jet  & Truth                & 0.2581 & 0.0060 & 0.0274 \\
                & \textsc{Comrade}     & 0.1907 & 0.0088 & 0.0756 \\
                & \textsc{Themis}      & 0.1802 & 0.0169 & 0.0799 \\
                & \textsc{Resolve}     & 0.2129 & 0.0286 & 0.1015 \\
                & \textsc{eht-imaging} & 0.1900 & 0.0520 & 0.1333 \\
\hline
low-jet  & Truth                & 0.1546 & 0.0053 & 0.0330 \\
                & \textsc{Comrade}     & 0.1062 & 0.0079 & 0.0715 \\
                & \textsc{Themis}      & 0.0976 & 0.0129 & 0.0462 \\
                & \textsc{Resolve}     & 0.1234 & 0.0349 & 0.0909 \\
                & \textsc{eht-imaging} & 0.1145 & 0.0443 & 0.0623 \\
\hline\hline
\end{tabular}
\textbf{Notes.} The jet $F_{\rm J}$, counter-jet $F_{\rm CJ}$, and outside $F_{\rm out}$ regions are defined on the truth image shown in Fig.~\ref{fig:2022jetrecovery}.
\end{table}

\begin{figure*}[h]
\centering
 \includegraphics[width=0.9\linewidth]{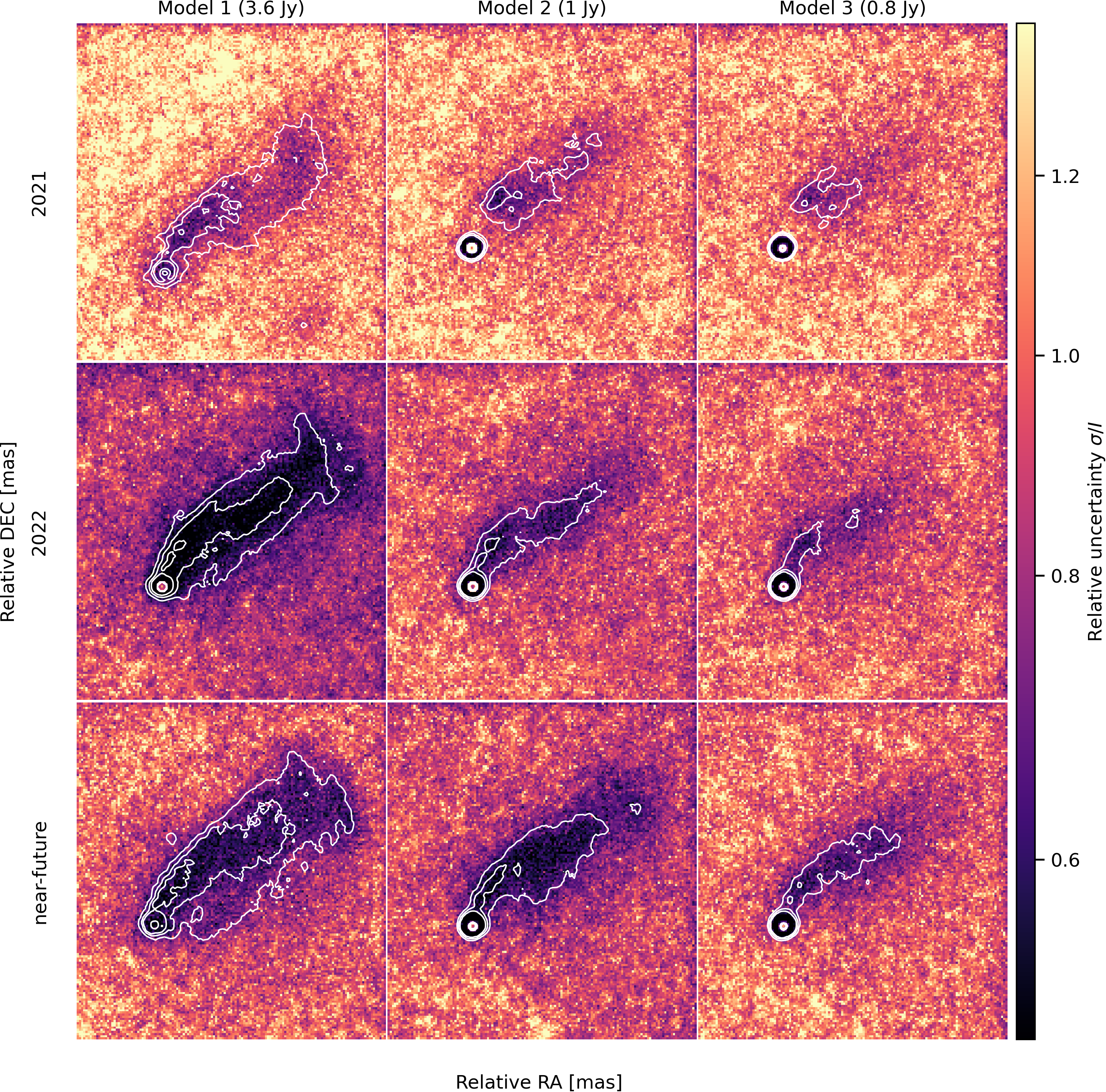}
\caption{Relative uncertainty maps, defined as $\sigma/I$, derived from the \textsc{Comrade} posterior distributions for the three models and observing campaigns. All maps are shown using the same field of view and colour scale. Darker regions correspond to lower relative uncertainty, while brighter regions indicate higher relative uncertainty. White contours indicate intensity levels increasing by factors of 3, with the lowest displayed contour at $27\,\sigma$ .} \label{fig:uncertainitymap}
\end{figure*}

\subsection{Array performance}
After establishing the lower limit for jet detection, we investigate how the recovery of the input models changes across array configurations. As mentioned above, we adopt \textsc{Comrade} as the reference imaging pipeline, as it provides reconstructions that are representative for both jet and counter-jet structures and exhibits the clearest trends across the different array configurations. This choice allows us to investigate how the quality of the reconstruction evolves across the considered observing epochs. 
Fig.~\ref{fig:uncertainitymap} presents the relative uncertainty maps derived from the \textsc{Comrade} posterior distributions, defined as the ratio between the posterior standard deviation, $\sigma$, and the posterior mean intensity, $I$, for the three input models and for the 2021, 2022, and near-future arrays. For each configuration, the three columns correspond to reconstructions assuming total flux densities of 3.6~Jy, 1~Jy, and 0.8~Jy, respectively. Darker regions correspond to lower relative uncertainty (i.e., higher posterior confidence), while brighter regions correspond to higher uncertainty.
In all cases, the relative uncertainty is lowest in regions close to the ring and the jet base and increases toward fainter emission and larger angular separations from the ring. This behavior is expected, as the relative uncertainty is inversely related to the local signal-to-noise ratio of the reconstructed emission. Along the jet direction, the emission is recovered with higher confidence near the jet base, while the uncertainty increases progressively toward the more extended regions of the jet.
The figure shows a clear improvement in the reconstruction of the dimmest input models (1~Jy and 0.8~Jy) with increasing array capability. For the 2021 configuration, the relative uncertainty is overall higher, and in the faintest models the jet-related contours appear detached from the main jet structure. In contrast, the near-future configuration exhibits more extended regions of reduced fractional uncertainty, with contours that remain more closely connected to the jet morphology even in the lowest flux density case.
The 2022 epoch shows generally good performance and in particular for the highest-flux model displays darker and more extended regions of low fractional uncertainty along the jet, indicating a robust reconstruction.
Even if the near-future configuration further improves the posterior confidence of the reconstructed jet, and both \textsc{Comrade} and \textsc{Themis} show improved reconstructions compared to 2022 (see Appendix~\ref{sec:2021&nearfuture}, Fig.~\ref{fig:nearfuture}), this progression is not seen across all imaging pipelines.
We attribute this behavior to the near-future synthetic dataset, which includes the introduction of KVN stations. As these stations are new to the array, their inclusion introduces additional calibration complexity, particularly in the presence of poor atmospheric conditions (e.g., elevated precipitable water vapour). These effects are reproduced in \textsc{Symba} and required additional flagging of scans with low detection significance, particularly on the longest baselines. 
Therefore, we interpret the near-future results conservatively, with some simulation parameters intentionally chosen to be conservative rather than fully realistic, given that the instrumental properties of the new stations (e.g., station sensitivity) and weather conditions are not yet characterized as well as in the 2021 and 2022 configurations.

We emphasize that the relative uncertainty maps trace where the posterior distribution is most tightly constrained and therefore indicate the confidence of the reconstruction. They do not directly quantify how closely the reconstructed emission matches the truth within the jet and counter-jet regions. 
This comparison is instead provided by the recovery ratios discussed in Sect.~\ref{sec:2022recovery} and listed in Tab.~\ref{tab:flux_recovery_2022}. These metrics show how the reconstructed emission is distributed among the ring, jet, and surrounding regions, and can differ even when the overall morphology is well constrained.
Taken together, the two analyses indicate that the 2022 array configuration provides the most balanced combination of robust jet recovery and consistency across imaging pipelines, while the near-future configuration yields the highest reconstruction confidence for the faintest jet structures.

\section{Conclusions}
\label{Sec:conclusions}

This work investigates the detectability of extended jet base emission in \m87 at EHT frequencies using synthetic observations generated with the \textsc{Symba} pipeline, which reproduces realistic EHT observing conditions, noise properties, and calibration effects. By varying the intrinsic jet brightness and considering different imaging approaches and array configurations representative of the 2021, 2022, and near-future observing epochs, we assess how jet-related emission is recovered under realistic interferometric conditions.
Recent analyses of the 2021 EHT observations have provided the first constraints at 230~GHz on faint, asymmetric emission on intermediate angular scales near \m87, suggesting the presence of weak jet-related structure close to the ring \citep{M872025, Saurabh2026, Georgiev2026}. These constraints were enabled thanks to the participation of NOEMA and KP, which improved the array’s sensitivity to diffuse emission compared to earlier EHT campaigns. Motivated by these results, we adopt the 2021 configuration as the starting case for investigating jet detectability.

Using both RML and Bayesian imaging methods, we find that jet emission can still be recovered in the faintest model considered here, which has a total flux density of 0.8~Jy. At this level, all imaging pipelines reconstruct a jet-like component, although fainter emission would be expected to become increasingly sensitive to noise and method-dependent effects.
Blind imaging tests confirm that, in this faintest case, recovering the jet remains challenging and is primarily limited by signal-to-noise. In these tests, the jet orientation was not disclosed to the imagers, ensuring that any recovered jet structure arises from the data rather than from prior assumptions about its direction.
The intrinsic counter-jet contribution is negligible in the mid- and low-jet models, such that even small amounts of misplaced emission can lead to large apparent counter-jet recovery ratios. This highlights the need for caution when interpreting counter-jet recovery metrics, particularly when the signal is very weak.

To explore how array improvements affect jet detectability, we then consider the 2022 configuration, which restores the LMT and introduces the first closed triangle of intermediate baselines (LMT–SMT–KP). This configuration provides the most balanced performance among those considered here, combining robust jet recovery ratios with consistent behavior across imaging pipelines.
Finally, we examine the near-future array configuration, which includes the additional KVN stations and provides a more densely sampled $(u,v)$ plane. In this case, the \textsc{Comrade} uncertainty maps show improved posterior confidence for faint jet structures, although these gains are not uniformly realized across all imaging methods, likely reflecting increased calibration uncertainties associated with the introduction of new stations.

Our simulations also place constraints on the brightness of the horizon-scale jet. 
In models with total flux densities of 0.8–1~Jy, the jet produces detectable signatures in the synthetic 2021 EHT data. The absence of similarly strong features in the real observations \citep{M872025} suggests that the jet emission on these scales is likely fainter, consistent with the limits derived by \citet{Saurabh2026}.

A caveat of this work is that the adopted jet models do not capture the ridge-brightened, knot-dominated morphology seen in existing 86 GHz images of M87. The present analysis should therefore be interpreted primarily as a study of diffuse inner-jet detectability and is likely conservative with respect to flux detection. If the true 230 GHz morphology is instead dominated by compact bright knots, such features could remain detectable even for lower total jet flux densities than in the models explored here. More realistic morphologies, including recent ring-plus-edge-brightened parabolic jet models developed to reproduce the GMVA+ALMA observations \citep{Gomez2026}, anisotropic nonthermal-electron models for limb-brightened jets \citep{Tsunetoe2025}, and slow-light radiative-transfer calculations \citep{Tsunetoe2026}, could be tested within the same simulation framework in future work.

We conclude that the improved intermediate-baseline coverage introduced in the 2022 EHT observation provides a stronger test of inner-jet emission on the relevant angular scales than was possible in 2021. If the inner jet contributes a substantial fraction of the unresolved compact flux density, it should therefore become more robustly detectable in post-2021 observations.

Upcoming analyses of the 2022 data will provide a direct test of whether the inner jet is responsible for this missing flux density in \m87. 
Future stations such as those in the Canary Islands and Namibia will further enhance European $(u,v)$ coverage and provide additional intermediate baselines with PV and NOEMA, improving phase information on large angular scales and enabling a more robust localization of the jet emission.
Moreover, future observations spanning multiple frequencies, epochs, and stations are expected to further improve the detectability of the jet base emission in \m87.

\begin{acknowledgements}
We thank Boris Georgiev for useful discussions and the anonymous EHT Collaboration internal reviewer for helpful comments. 
\end{acknowledgements}

\bibliographystyle{aa.bst} 
\bibliography{biblio.bib} 

\clearpage
\appendix
\section{Additional tests}
\subsection{The 2021 and near-future configurations}
\label{sec:2021&nearfuture}

\begin{figure*}[p]
    \centering
    \includegraphics[width=\linewidth]{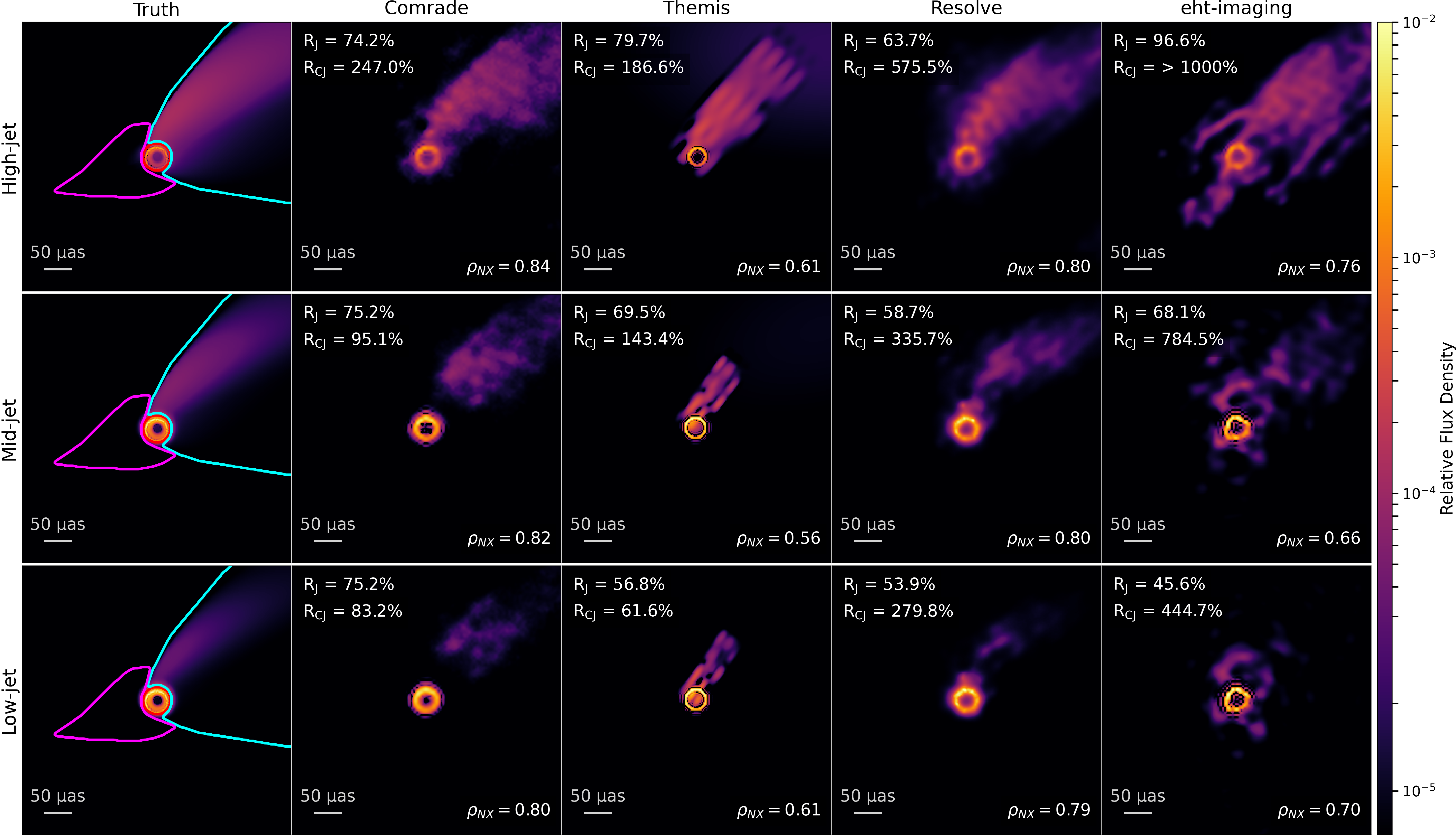}
    \caption{Jet recovery for the 2021 array configuration.
The layout and annotations follow Fig.~\ref{fig:2022jetrecovery}.
All images are aligned to the truth and normalized to 1~Jy.}
\label{fig:2021comparison}
\end{figure*}

\begin{figure*}[p]
    \centering
    \includegraphics[width=\linewidth]{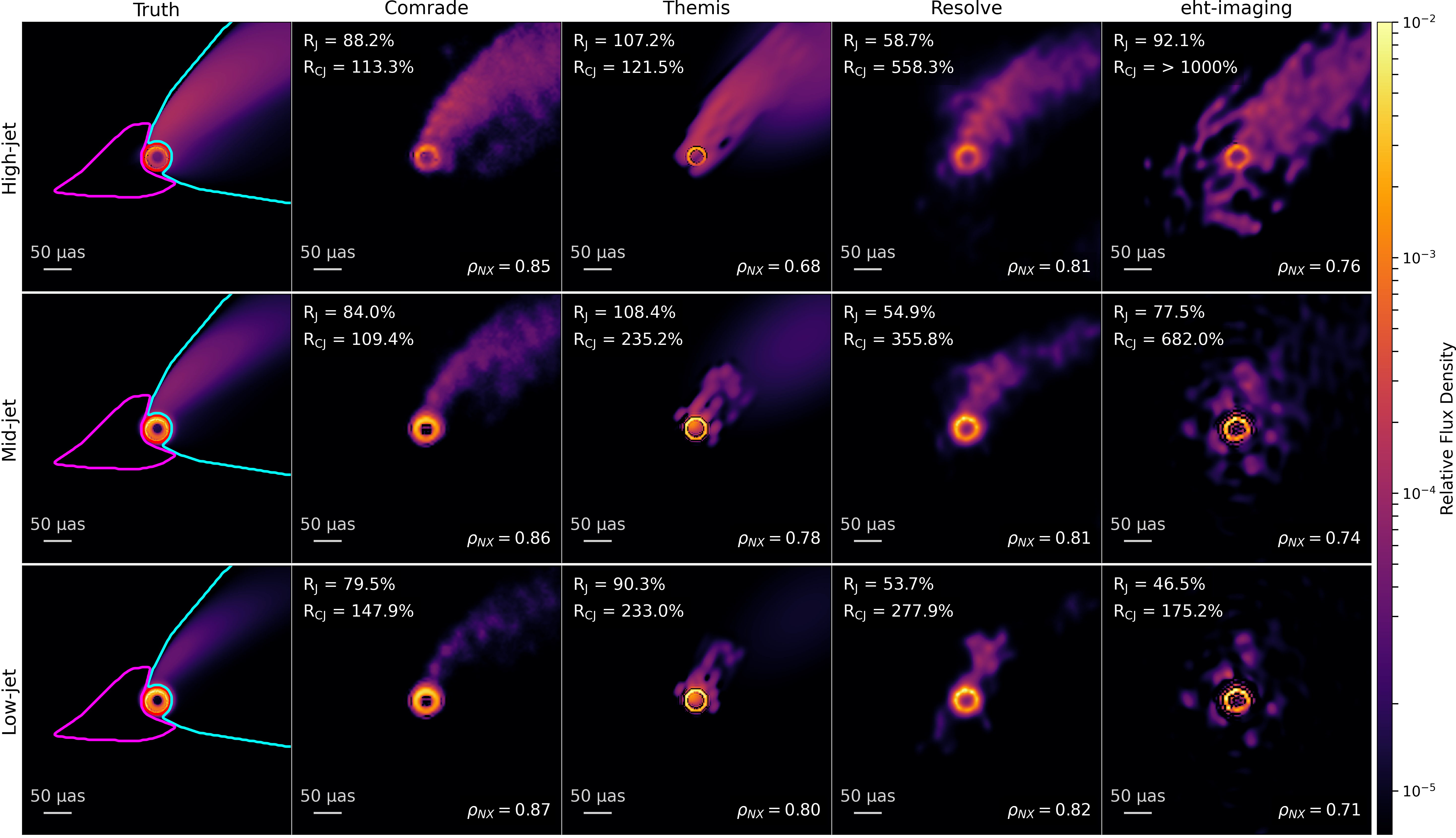}
    \caption{Jet recovery for the near-future array configuration.
The layout and annotations follow Fig.~\ref{fig:2022jetrecovery}.
All images are aligned to the truth and normalized to 1~Jy. }
\label{fig:nearfuture}
\end{figure*}

In this section, we extend the jet recovery analysis presented in Fig.~\ref{fig:2022jetrecovery} to the 2021 and near-future EHT array configurations. The same analysis procedure described in Sect.~\ref{sec:2022recovery} is applied: all images are aligned to the center of the truth ring, normalized to a total flux density of 1~Jy, and evaluated using fixed ring, jet and counter-jet regions.
Fig.~\ref{fig:2021comparison} and \ref{fig:nearfuture} show the reconstructed images together with the corresponding jet and counter-jet recovery ratios ($R_{\rm J}$ and $R_{\rm CJ}$), expressed as percentages relative to the truth model, and the normalized cross-correlation coefficient $\rho_{\rm NX}$.
The overall behavior is consistent with that found for the 2022 configuration. In particular, the recovered jet ratio $R_{\rm J}$ decreases systematically from the high-jet to the low-jet model for all imaging pipelines.
The counter-jet ratios $R_{\rm CJ}$ are generally much larger and more variable. As discussed in Appendix~\ref{sec:counterjet}, the intrinsic counter-jet emission in the mid- and low-jet models is extremely weak, such that even small amounts of reconstructed emission in this region can produce large $R_{\rm CJ}$ values (see also Tab.~\ref{tab:fluxes_2021} and Tab.~\ref{tab:fluxes_nearfuture}).
The 2021 array generally provides the weakest jet recovery across the imaging pipelines. The comparison between the 2022 and near-future configurations is less uniform. In particular, \textsc{Comrade} and \textsc{Themis} tend to show slightly higher recovered jet fractions for the near-future array, whereas this trend is not consistently observed for \textsc{Resolve} and \textsc{eht-imaging}.
As discussed in Sect.~\ref{sec:2022recovery}, these differences should be interpreted with caution. First, the near-future synthetic dataset was generated using conservative assumptions regarding station performance and atmospheric conditions, which can limit improvements observed. Second, the analysis presented in Fig.~\ref{fig:2021comparison} and Fig.~\ref{fig:nearfuture} is based on images normalized to the same total flux density, which suppresses differences between observing campaigns. As a result, the recovered flux density fractions are not necessarily the most sensitive diagnostic for comparing array performance across years.
For this reason, in the main analysis we complement the flux density recovery metrics with the relative uncertainty maps derived from the \textsc{Comrade} posterior distributions (Fig.~\ref{fig:uncertainitymap}), which provide a more robust diagnostic of the reconstruction confidence across the different array configurations.

\subsection{Blind tests}
\label{Blind_test}
\begin{figure*}[h]
    \centering
    \includegraphics[width=\linewidth]{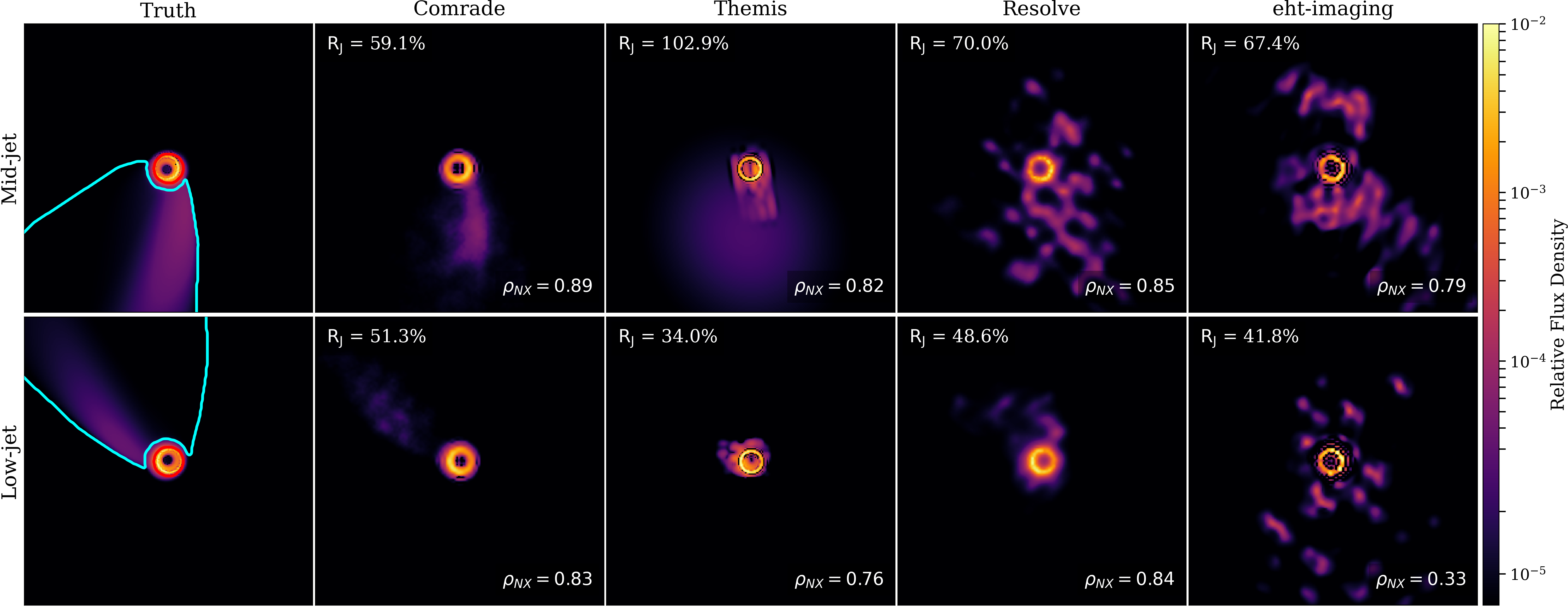}
    \caption{Blind tests using the 2022 configuration and the mid- and low-jet models. The layout and annotations follow Fig.~\ref{fig:2022jetrecovery}. All images are aligned to the truth and normalized to 1~Jy.}
    \label{fig:blindtest}
\end{figure*}

To assess whether the recovery of faint jet emission is driven by the data or instead influenced by prior knowledge of the jet direction, we performed a set of blind imaging tests.
Fig.~\ref{fig:blindtest} shows the reconstructions obtained for two representative blind tests, generated using the mid-jet and low-jet models with the 2022 array configuration. In both cases, the intrinsic jet orientation was rotated and this information was not disclosed to the imagers. All other aspects of the analysis were kept identical to the main jet recovery study, including the array configuration, alignment procedure, region definitions, and flux density normalization.
The recovered jet flux density ratio $R_{\rm J}$ and the normalized cross-correlation coefficient $\rho_{\rm NX}$ are reported in each panel.
For the mid-jet model (first row), all imaging pipelines recover a detectable jet component, with recovered jet ratios typically in the range $R_{\rm J}\sim 50 -100\%$. Although the reconstructed jet morphology is generally more diffuse than in the non-blind case, it broadly follows the true jet direction, indicating that the recovery is not driven by prior information.
The reconstruction becomes significantly more challenging for the low-jet model (second row). In this case, most imaging pipelines struggle to recover a coherent jet morphology, with fragmented or weak jet-like features and reduced $\rho_{\rm NX}$ values. Among the tested methods, \textsc{Comrade} most consistently recovers a jet morphology that resembles the truth model, while the other imaging pipelines show larger deviations.
These blind tests confirm that jet recovery at low intrinsic jet flux density is primarily limited by signal-to-noise, leading to increasingly fragmented and method-dependent reconstructions. This result further supports our choice of 0.8 Jy total flux density as a practical threshold for reliable jet detectability at this observing frequency, since models with fainter jets would likely yield increasingly unreliable and method-dependent jet features. 


\begin{table}
\centering
\caption{Integrated flux densities for the 2021 EHT array.}
\label{tab:fluxes_2021}
\begin{tabular}{llccc}
\hline\hline
Model & Pipeline & $F_{\rm J}$ [Jy] & $F_{\rm CJ}$ [Jy] & $F_{\rm out}$ [Jy] \\
\hline
high-jet & Truth                & 0.5347 & 0.0075 & 0.0098 \\
                & \textsc{Comrade}     & 0.4102 & 0.0174 & 0.0736 \\
                & \textsc{Themis}      & 0.5186 & 0.0102 & 0.1322 \\
                & \textsc{Resolve}     & 0.3919 & 0.0322 & 0.1659 \\
                & \textsc{eht-imaging} & 0.4683 & 0.0546 & 0.0934 \\
\hline
mid-jet  & Truth                & 0.2581 & 0.0060 & 0.0274 \\
                & \textsc{Comrade}     & 0.1959 & 0.0076 & 0.0585 \\
                & \textsc{Themis}      & 0.1946 & 0.0110 & 0.0829 \\
                & \textsc{Resolve}     & 0.1802 & 0.0223 & 0.1046 \\
                & \textsc{eht-imaging} & 0.1593 & 0.0394 & 0.0999 \\
\hline
low-jet  & Truth                & 0.1546 & 0.0053 & 0.0330 \\
                & \textsc{Comrade}     & 0.1118 & 0.0058 & 0.0644 \\
                & \textsc{Themis}      & 0.1146 & 0.0070 & 0.0562 \\
                & \textsc{Resolve}     & 0.0941 & 0.0176 & 0.0836 \\
                & \textsc{eht-imaging} & 0.0724 & 0.0309 & 0.0660 \\
\hline\hline
\end{tabular}

\noindent
\end{table}

\begin{table}
\centering
\caption{Integrated flux densities for the near-future EHT array.}
\label{tab:fluxes_nearfuture}
\begin{tabular}{llccc}
\hline\hline
Model & Pipeline & $F_{\rm J}$ [Jy] & $F_{\rm CJ}$ [Jy] & $F_{\rm out}$ [Jy] \\
\hline
high-jet & Truth                & 0.5347 & 0.0075 & 0.0098 \\
                & \textsc{Comrade}     & 0.4690 & 0.0080 & 0.0571 \\
                & \textsc{Themis}      & 0.5942 & 0.0102 & 0.0540 \\
                & \textsc{Resolve}     & 0.3497 & 0.0311 & 0.1864 \\
                & \textsc{eht-imaging} & 0.4522 & 0.0615 & 0.1389 \\
\hline
mid-jet  & Truth                & 0.2581 & 0.0060 & 0.0274 \\
                & \textsc{Comrade}     & 0.2138 & 0.0078 & 0.0628 \\
                & \textsc{Themis}      & 0.2391 & 0.0110 & 0.0522 \\
                & \textsc{Resolve}     & 0.1711 & 0.0240 & 0.1130 \\
                & \textsc{eht-imaging} & 0.1544 & 0.0437 & 0.1144 \\
\hline
low-jet  & Truth                & 0.1546 & 0.0053 & 0.0330 \\
                & \textsc{Comrade}     & 0.1214 & 0.0077 & 0.0619 \\
                & \textsc{Themis}      & 0.1419 & 0.0089 & 0.0356 \\
                & \textsc{Resolve}     & 0.1027 & 0.0229 & 0.0857 \\
                & \textsc{eht-imaging} & 0.0725 & 0.0357 & 0.0775 \\
\hline\hline
\end{tabular}
\end{table}

\begin{figure*}[h]
    \centering
    \includegraphics[width=\linewidth]{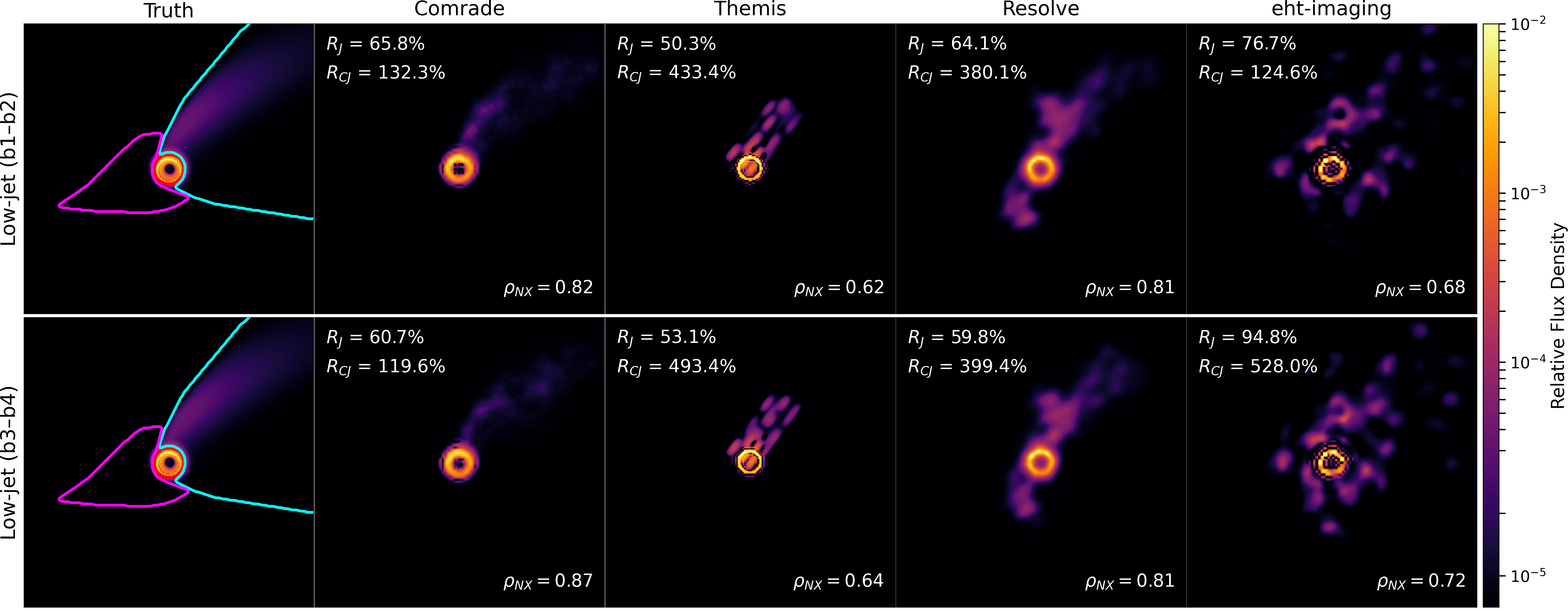}
    \caption{Bands consistency test using the 2022 configuration and the low-jet model. The layout and annotations follow Fig.~\ref{fig:2022jetrecovery}. All images are aligned to the truth and normalized to 1~Jy. }
    \label{fig:bandconsistency}
\end{figure*}

\subsection{Bands consistency}
\label{sec:Band_consistency}

To increase the sensitivity of the array, we combine EHT bands 1–2 and 3–4, respectively. In this section, we test the consistency of our results by repeating the jet recovery analysis using the independent $b3$–$b4$ band pair. We find that the jet detection limit remains unchanged when using the $b3$–$b4$ dataset. Fig.~\ref{fig:bandconsistency}  shows the reconstructions for the low-jet model using the 2022 array configuration, where the first and second rows correspond to $b1$-$b2$ and $b3$-$b4$, respectively. The recovered jet flux density ratios and $\rho_{\rm NX}$  are generally consistent with those obtained for the $b1$–$b2$ combination, confirming a good agreement between the two band pairs.
This consistency indicates that our conclusions on jet detectability are independent of the choice of band combination. While the current study uses band combinations at 230~GHz ($b1$–$b2$  and $b3$–$b4$), ongoing efforts to extend EHT observations to higher frequencies (e.g., 345~GHz) and wider bandwidths are expected to provide even better constraints on jet morphology and flux density recovery.

\section{Counter-jet detectability}
\label{sec:counterjet}
\begin{figure*}[h]
    \centering
    \includegraphics[width=\linewidth]{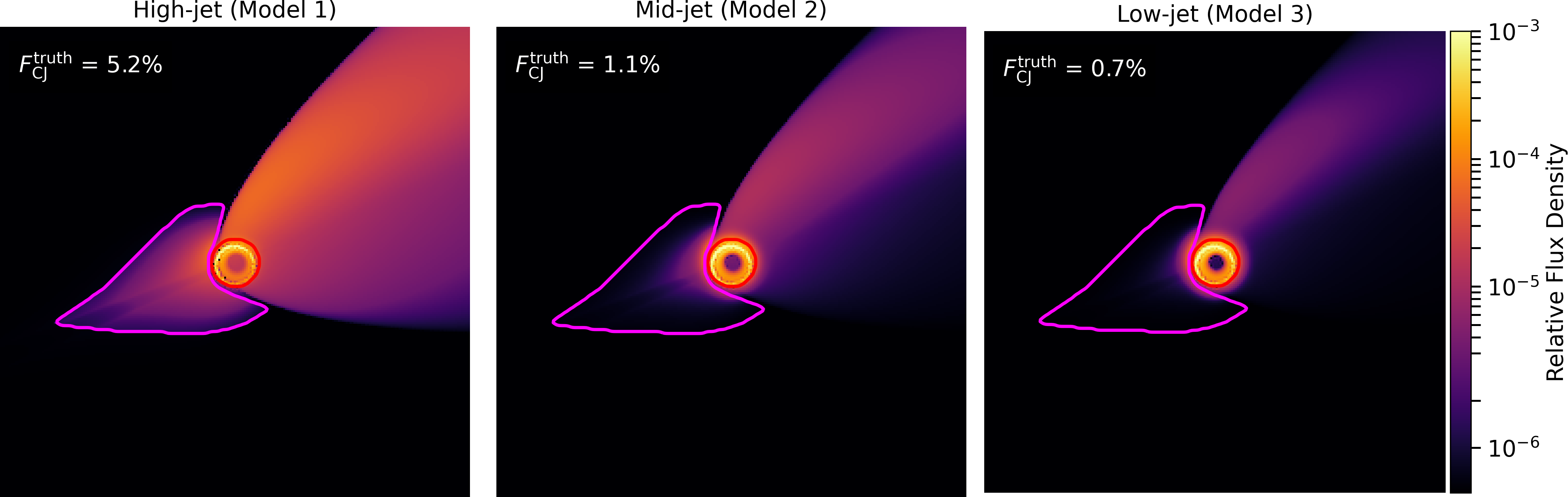}
    \caption{Counter-jet emission in the truth models. The reported counter-jet flux density fraction is computed relative to the ring emission (red contour).}
    \label{fig:CJ}
\end{figure*}

Fig.~\ref{fig:CJ} shows the three truth models (high-, mid-, and low-jet)  displayed with identical intensity scaling. The red contour marks the ring region used for normalization, and the magenta contour marks the counter-jet (CJ) region. In each panel, we report the counter-jet flux density fraction relative to the ring,
\begin{equation}
F_{\rm CJ}^{\rm truth}
=
\frac{\sum_{p \in \rm CJ} I^{\rm truth}(p)}
       {\sum_{p \in \rm ring} I^{\rm truth}(p)},
\end{equation}
where $p$ indexes image pixels and negative pixel values are excluded from the sums. We find $F_{\rm CJ}^{\rm truth} = 5.2\%$, $1.1\%$, and $0.7\%$ for the high-, mid-, and low-jet models, respectively, indicating that the counter-jet contribution rapidly becomes negligible. In contrast, the jet flux density fractions remain significant, with $F_{\rm J}^{\rm truth}$ of approximately $71\%$ and $33\%$ for the mid- and low-jet models, highlighting the strong asymmetry between jet and counter-jet emission. As a consequence, even modest reconstructed emission in the counter-jet region can translate into large relative values. This explains why some imaging pipelines yield elevated counter-jet ratios, despite the intrinsically weak counter-jet signal. More generally, this highlights the need for caution when interpreting counter-jet flux density measurements, including in analyses of real EHT data.
\section{Closure phases}
\begin{figure*}[h]
    \centering
    \includegraphics[width=\linewidth]{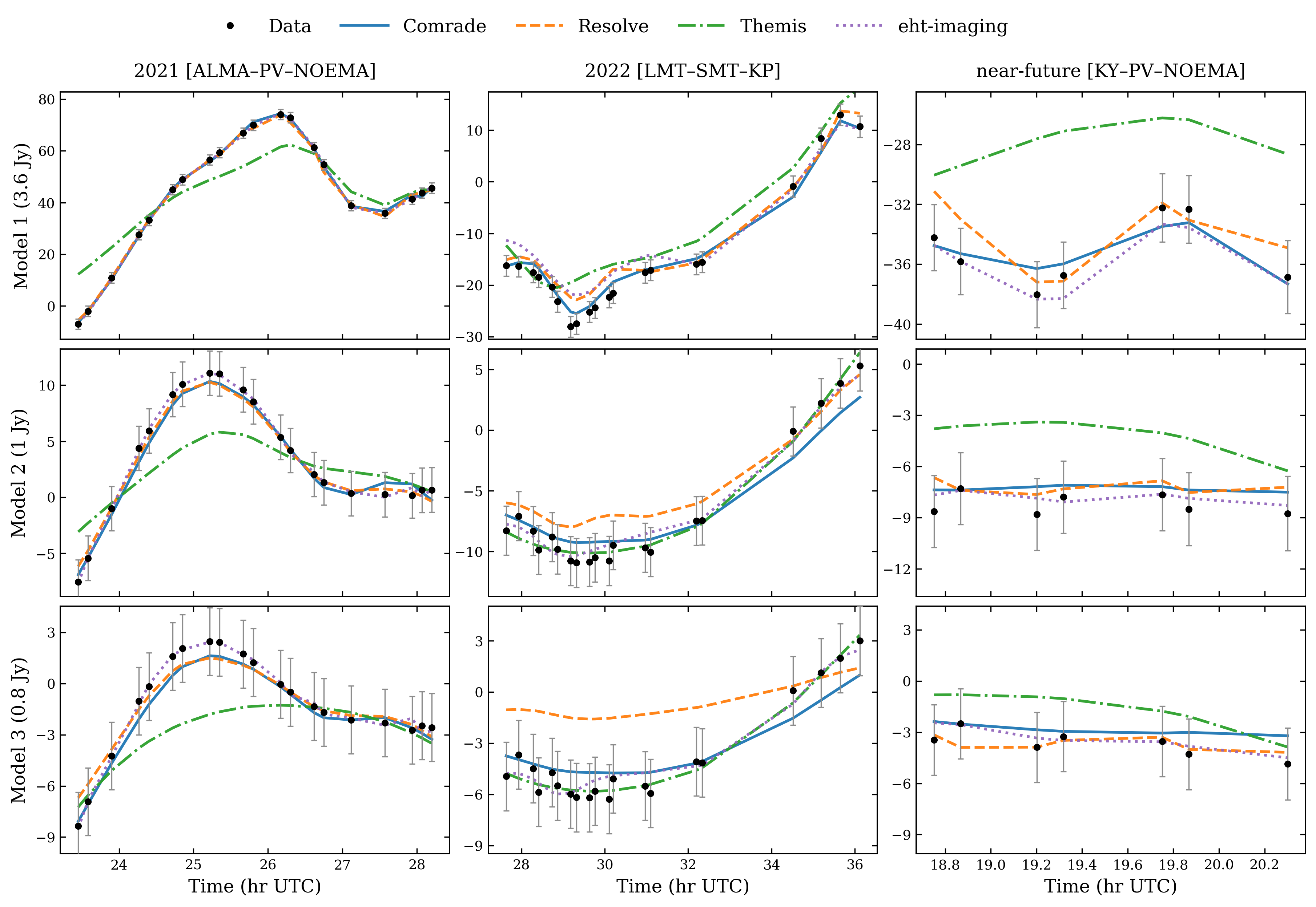}
    \caption{Representative closure phase triangles considered in this work. Rows correspond to the three input models with progressively fainter intrinsic jet emission, while columns show the 2021, 2022, and near-future EHT configurations.}
    \label{fig:closures}
\end{figure*}

Closure phases are interferometric observables formed by summing phases around closed baseline triangles. They are independent of station-based phase errors and therefore provide a robust probe of source asymmetry and extended emission in EHT observations. Closure phases have been extensively used to study deviations from azimuthal symmetry in \m87, including signatures associated with jet base emission \citep[e.g.][]{Chael2016, M87P1,Saurabh2026, Georgiev2026}.
In this work, closure phases are used as a consistency check between the synthetic data and the reconstructed images for models with different intrinsic jet brightness. Fig.~\ref{fig:closures} presents representative closure triangles for the synthetic datasets considered here, covering the 2021, 2022, and near-future array configurations and the three input models with progressively fainter jets.
For each observing year, a single representative closure triangle is shown, chosen to reflect the evolution of the array configuration and the introduction of new stations. Specifically, we show the ALMA-PV-NOEMA triangle for 2021, the LMT-SMT-KP triangle for 2022, and the KY-PV-NOEMA triangle for the near-future case.
Recent closure phase studies of \m87 with the 2021 array have shown that specific high-sensitivity triangles, such as ALMA–SMT–KP and ALMA-PV-NOEMA, are interpreted as signatures of faint asymmetric emission on intermediate spatial scales \citep{Saurabh2026, Georgiev2026}. In our synthetic data, the same triangles show similar closure phase behavior.
As discussed above, the LMT–SMT–KP triangle introduced in the 2022 configuration represents the first closed triangle of intermediate baselines probing compact and jet base angular scales in \m87. Finally, the KY–PV–NOEMA triangle was chosen as a representative example of increased redundancy in the expanded near-future array. Fig.~\ref{fig:closures} also shows that closure phase deviations increase with jet brightness, with the strongest jet model producing the largest deviations.

\section{\m87 EHT coverage}

\label{Sec:2017}
\begin{figure}[h]
   \centering
  \includegraphics[width=\linewidth]{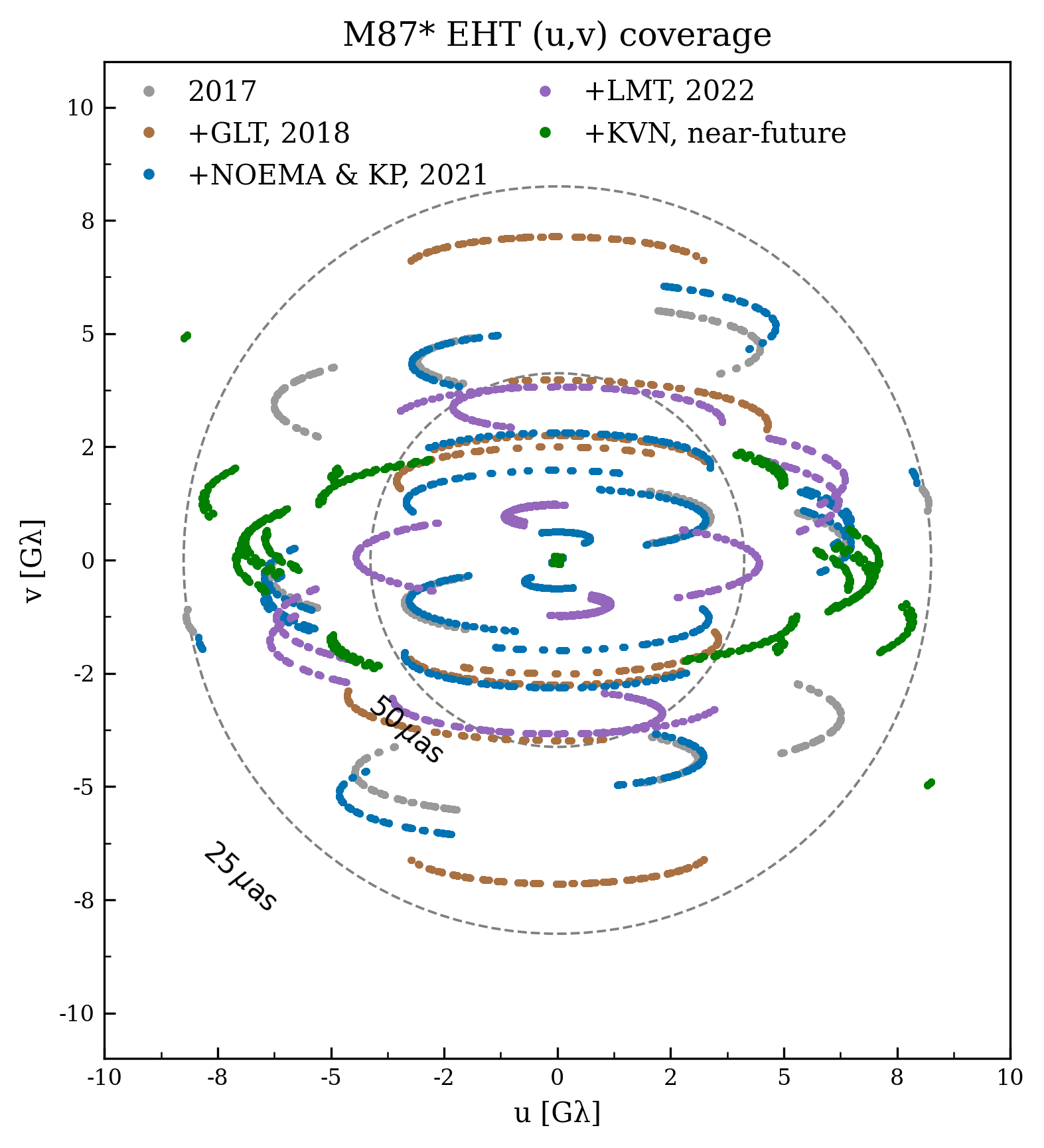}
   \caption{Projected $(u,v)$ coverage of \m87 at 230~GHz, illustrating the incremental expansion of the EHT array since 2017. Gray points show the baseline coverage of the 2017 configuration, while colored points highlight additional baselines introduced in subsequent observations.} \label{fig:uvplots_2017}
\end{figure}

As explained in the main text (Sect.~\ref{Sec:array}), the EHT array has undergone continuous upgrades since the 2017 observations that produced the first image of \m87. In Fig.~\ref{fig:uvplots_2017}, we illustrate the evolution of the \m87 $(u,v)$ coverage over the years, highlighting in different colors the additional telescopes incorporated into the array. The 2017 coverage is shown as gray points and serves as a reference baseline. The inclusion of GLT (brown points) represents a particularly important upgrade for \m87 observations, as it provides long north–south baselines that significantly improve angular resolution \citep{M872018}. Subsequent expansions of the array in 2021, 2022, and the near-future are shown using the same color scheme adopted in Fig.~\ref{fig:uvplots} of the main text. As mentioned above, these later additions mostly contribute to shorter baselines, leading to a progressively denser filling of the central $(u,v)$ plane. However, the short baselines ($\lesssim 1\,\mathrm{G}\lambda$) remain relatively sparse, which continues to limit the sensitivity to diffuse jet emission. Future expansions of the array will therefore be important for improving the detection of faint jet structures.

\section{Imaging pipeline setup}
\label{sec:methods details}
\subsection{\textsc{Comrade}}

\textsc{Comrade} is a differentiable Bayesian modeling package designed to model generic VLBI data, including sparse EHT data, written in julia.  The latest version of \textsc{Comrade} tackles VLBI image reconstruction using hierarchical Bayesian priors, producing posteriors on the direct imaging parameters, model hyper-parameters like the raster point correlation coefficients, and instrument parameters like the telescope gain amplitudes and phases.  The particular model used in this work is similar to the parameterized raster model described in \citet{Broderick2020_raster} and \citet{Tiede2022}, and discussed in detail in \citet{Tiede2026_HIBI}. The image model is parameterized as a stochastic process consisting of a rasterized mean image structure with some correlation.  The prior image for this paper was assumed to be a ring, with a Gaussian mean random field applied as "noise" to modulate the emission.  The instrument parameters (gain amplitudes and phases) were fit simultaneously with the image parameters.  The reconstruction region was restricted to an FOV of 1 mas, so to avoid artifacts from the unmodeled intrasite baselines (providing image information at $\sim$as scales) we removed visibilities below 0.05 G$\lambda$ and only fit the remainder.  We fit the scan-averaged complex visibility data with an additional $2\%$ fractional systematic uncertainty meant to reflect unresolved calibration artifacts like leakage and/or coherence loss \citep{M87P3,M872018}.

\subsection{\textsc{Themis}}

\textsc{Themis} is a modular Bayesian parameter estimation framework developed to directly compare parameterized models to VLBI data products produced by the EHT, primarily written in C++ and designed to be used on large-scale HPC systems.  The particular parameterized model used in the work is composed of an analytically defined geometric crescent to model the bright ring, an adaptive spline raster grid to model arbitrary emission around the ring, and a large scale asymmetric Gaussian to model the emission at the largest scales.  Posteriors on each model component are sampled using an adaptive parallel tempered Hamiltonian Monte Carlo sampler. The instrument gains for each telescope are recovered via optimization in each model evaluation step.  This particular model has been used in the analysis of the 2017 \citep{Broderick2022_PhotonRing} and 2018 \m87 \citep{M872018} data.  The model used in this work is identical to the "Hybrid Themage" model described in Section 6.2.1 of \citet{M872018}, except that this paper uses a 7x7 raster instead of a 5x5 raster, since the near-future coverage supports additional model complexity.  The model priors used here are the same as those described in Appendix J of \citet{M872018}.  The synthetic data was also processed in the same way as done in \citet{M872018}, in that we fit to scan averaged complex visibilities, with $2\%$ additional fractional systematic uncertainty in quadrature with the thermal uncertainties, similar to the \textsc{Comrade} reconstructions.

\subsection{\textsc{Resolve}}

\textsc{Resolve} is a Bayesian imaging framework for radio interferometry that has been used in VLBI imaging of both \m87 and \sgra \citep{Arras2022,Resolve2023,Kim2024,Kim2025}.
\textsc{Resolve} objective is to infer the posterior distribution of the sky brightness $I$, given the interferometric visibility data $V$, by using the negative logarithm of Bayes' theorem. This results in $H(I|V) = H(V|I) + H(I) - H(V)$, where $H$ denotes the negative logarithmic probability, also known as the information Hamiltonian. $H(V|I)$ represents the negative log-likelihood and enforces data fidelity, while $H(I)$ encodes prior information about the image structure. The evidence term, $H(V)$, is independent of $I$ and can therefore be neglected during optimization.
We model the brightness distribution and the instrumental calibration parameters as Gaussian processes with a correlation kernel that is inferred jointly during the optimization. This approach enables adaptive regularization directly from the data.
For initialization, the reconstruction starts with a Gaussian brightness distribution of $30 \, \mu \rm{as} \times 30 \, \mu \rm{as}$, where we assumed 0.0 systematic noise. We utilized a sky model covering an FOV of 1050 by 1050 $\mu \rm{as}$ using 500 by 500 pixels. During inference, 16 pairs of antithetic samples are used per iteration to efficiently estimate the latent parameters and their variance.

\subsection{\textsc{eht-imaging}}

Regularized Maximum Likelihood (RML) imaging reconstructs an image by combining constraints from the data with prior assumptions about the source structure, such as smoothness, sparsity, or entropy. This is formulated as an optimization problem in which a weighted sum of data fidelity terms and regularization functionals is minimized:
\begin{align}
\hat{I} = \arg\min_I \left( \sum_d \alpha_d \chi^2_d(I) + \sum_x \beta_x R_x(I) \right).
\end{align}
The \textsc{eht-imaging} library provides a direct implementation of this framework. For a detailled overview of the standard regularization terms applied for analyses within the \textsc{eht-imaging} framework (and for RML methods in general), we refer to \citet{M87P4, M872025}, especially table C.1 in latter one.

The reconstruction depends on the choice of hyperparameters $\alpha_d$ and $\beta_x$ (which control the trade-off between agreement with the data and adherence to prior assumptions), as well as an iterative optimization strategy including blurring the guess solution, reinitializing and iterative self-calibration. In earlier work, the hyperparameters $\alpha_d$ and $\beta_x$ were explored through systematic surveys over a range of geometric source models \citep[e.g.][]{M87P4}. For the purpose of jet detection, we however noticed that heuristical pipeline choices (such as addition of systematic noise, how to handle the zero-baseline flux, iterative phase self-calibration and re-initialization of the imager) dominate the imaging performance over hyperparameter choices. 

In the present study therefore, we manually adjust pipeline heuristics and regularization weights to the best of our knowledge, and perform only a limited parameter exploration. For the initialization, we first fit a simple parametric model consisting of an asymmetric ring and an elliptical Gaussian. This is achieved via a grid search over ring diameter, asymmetry, and position angle, as well as Gaussian flux, size, and orientation. The resulting model is then used for self-calibration, after which the full \textsc{eht-imaging} reconstruction procedure is applied.

\section{Synthetic data generation}
\label{Sec:SYMBA}

The synthetic datasets used in this work were generated with \textsc{Symba} \citep{Roelofs2020}, which combines \textsc{MeqSilhouette} \citep{Natarajan2022} for realistic data corruption with the \textsc{rPICARD} pipeline \citep{Janssen2019} for calibration. For each configuration, \textsc{Symba} was driven by inputs reflecting the actual or expected observing conditions, so that the resulting $(u,v)$ coverage, noise properties, and residual calibration errors reproduce those of real EHT data as closely as possible.

For the 2021 configuration, we adopted the \m87 observing schedule, station list, and System Equivalent Flux Densities (SEFDs) of the 2021 EHT campaign \citep{M872025}.   Atmospheric corruptions, including opacity, sky temperature, and phase fluctuations, were computed by \textsc{MeqSilhouette} from the precipitable water vapour (PWV), ground temperature, and ground pressure recorded at each site during the campaign night.

For the 2022 configuration, the same procedure was followed using the corresponding schedule, station list, now including the LMT, SEFDs, and site-specific weather logs.

For the near-future configuration, the schedule was built by extending the 2022 schedule with additional scans on the KVN stations. For all stations except KY and PC, the 2022 weather logs and SEFDs were retained. For the new stations, we adopted representative seasonal PWV values and SEFDs based on comparable stations since the instrumental properties of KY and PC at 230\,GHz are not yet fully characterized. Low-S/N scans on the longest KVN baselines were flagged when their detection significance fell below the \textsc{rPICARD} fringe-detection threshold. A more realistic characterization of the KVN stations at 230\,GHz will be possible once dedicated observations become available and may improve the jet detection performance presented here.

\end{document}